\documentstyle[amssymb,aps,prl,epsfig,floats]{revtex}

\DeclareMathAlphabet{\mabold}{OT1}{cmr}{bx}{it}
\newcommand{\rb}{{\mabold{x}}}
\newcommand{\bqbot}{{\mabold{q}_{\bot}}}

\newcommand{\sbo}{{\mabold{s}}}

\newcommand{\beq}{\begin{equation}}
\newcommand{\eeq}{\end{equation}}

\newcommand{\scal}{{\cal S}}
\newcommand{\kcal}{{\cal K}}
\newcommand{\xbot}{{\mabold{x}}_{\bot}}
\newcommand{\dbot}{d_{\bot}}

\newcommand\mean[1]{\left\langle#1\right\rangle}
\newcommand{\RR}{{\mathbb R}}
\newcommand{\rl}{\leftrightarrow}

\begin{document}
\draft

%abrimos
\twocolumn[\hsize\textwidth\columnwidth\hsize\csname @twocolumnfalse\endcsname

\title{Internal Fluctuations Effects on Fisher Waves}

\author{Esteban Moro$^\dag$}

\address{Theoretical Physics, University of Oxford,
	 1 Keble Road OX1 3NP, United Kingdom}

\date{\today}

\maketitle

\begin{abstract}
We study the diffusion-limited reaction $A + A \rl A$ in various
spatial dimensions to observe the effect of internal fluctuations on
the interface between stable and unstable phases. We find that,
similar to what has been observed in $d=1$ dimensions, internal
fluctuations modify the mean-field behavior predictions for this
process, which is given by Fisher's reaction-diffusion equation.
In $d > 1$ the front displays local fluctuations
perpendicular to the direction of motion which, with a proper
definition of the interface, can be fully described within the
Kardar-Parisi-Zhang (KPZ) universality class. This clarifies the
apparent discrepancies with KPZ predictions reported recently.
\end{abstract}

% 5.40.-a Fluctuation phenomena, random processes, noise, and Brownian motion
% 5.70.Np Interface and surface thermodynamics
% 68.35.Rh Phase transitions and critical phenomena
% 68.35.Ct Interface structure and roughness
% 82.65.+r Surface and interface chemistry; heterogeneous catalysis at surfaces
% 81.10.Aj Theory and models of crystal growth; physics of crystal growth,..
% 64.60.Ht Dynamic critical phenomena
%y cerramos
]

Fluctuations in the macroscopic behavior of reaction-diffusion (RD)
systems could play an important role whether these fluctuations are
produced by the discrete nature of the elementary constituents ({\em
internal}) or by environmental random variations ({\em
external}). Often both type of fluctuations are neglected in the
theoretical treatment, thus describing the system by an action-mass
type equation (mean-field approximation). Fluctuations in RD systems can, for
instance, give rise to instabilities \cite{nature}, modify the
reaction front velocity
\cite{derrida,sancho1}, allow the system to reach new states absent in
the mean-field description
\cite{sancho}, or produce spatial correlations in the system which in
turn can dominate the macroscopic system behavior
\cite{cardy}. While these effects appear in different situations, 
there is a theoretical and experimental interest in the problem of
front propagation in RD systems. The simplest example is that of invasion of
an unstable phase by a stable one, which at the mean-field
approximation level is described by Fisher's equation
\cite{fisher}
\begin{equation}\label{fisher_equation}
\frac{\partial \rho}{\partial t} = D \nabla^2 \rho + k_1 \rho - k_2 \rho^2
\end{equation}
where $\rho(\rb,t)$ is the local concentration ($\rb \in \RR^d$)
characterizing the system.  Equation (\ref{fisher_equation}) arises in
the macroscopic description of many processes in physics, chemistry
and biology and is a generic model for reaction front propagation in
systems undergoing a transition from a marginally unstable ($\rho =
0$) to a stable ($\rho_{\mathrm{eq}} = k_1/k_2$) state.  Thus, for
initially segregated conditions, i.e. $\rho(\rb) = \rho_{\mathrm{eq}}$
for $x_{\|} \leq 0$ and $\rho(\rb) = 0$ for $x_{\|} > 0$ (with $\rb =
(\xbot,x_{\|})$, $\xbot \in \RR^{d_{\bot}}$, $x_{\|} \in
\RR$ and $d = \dbot + 1$), the solution of
(\ref{fisher_equation}) is a front invading the unstable phase and
propagating along $x_{\|}$ with a constant velocity $v \geq v_{min} = 2 (k_1
D)^{1/2}$ which is selected depending on the initial condition
according to the ``marginal stability criterion'' \cite{msc}.  At the
same time, the front broadens until it reaches a finite width $\xi
\simeq 8 (D/k_1)^{1/2}$.

The question of how faithfully continuum equation
(\ref{fisher_equation}) resembles the macroscopic front dynamics of
microscopic RD discrete systems has drawn a lot of
attention recently. In particular, much attention has been
devoted to microscopic stochastic models like $A+A \rl A$ or $A + B
\to 2A$, where $A$ and $B$ are active species. Discreteness of
those systems is responsible for fluctuations in $\rho(\rb,t)$ and
introduces an effective cutoff in the reaction-rates which modifies
the properties of the front.  Most of the studies have concentrated on
observing how the microscopic system approaches the macroscopic
behavior described by Eq.\ (\ref{fisher_equation}) in $d=1$ when $N
\to\infty$, where $N$ is the number of particles per site \cite{derrida}. 
Using van Kampen's system size expansion \cite{gardiner} or
field-theory techniques \cite{rmp}, the mesoscopic dynamics of the
microscopic Master equation can be expressed in terms of a Langevin
equation which, in the case of the $A + A \rl A$ scheme, reads
\cite{pechenik}
\beq\label{langevin}
\frac{\partial \rho}{\partial t} = D \nabla^2 \rho + k_1
\rho  - k_2 \rho^2 + \sqrt{(k_1 \rho -k_2 \rho^2)}\ \eta(\rb,t)
\eeq
where $\eta(\rb,t)$ is a uncorrelated white noise,
$\mean{\eta(\rb,t)}=0$, $\mean{\eta(\rb,t)\eta(\rb',t')} =
2 \hat N^{-1/2}\delta(\rb-\rb')\delta(t-t')$, and $\hat N$ is a reference
level of number of particles, which is proportional to $N$. In the
limit $N \to \infty$, the noise term in (\ref{langevin}) (which
reflects the fluctuations in the number of particles) vanishes and we
recover (\ref{fisher_equation}). In this limit, the effective cutoff
in the reaction term imposed by the discreteness seems to be the
leading contribution, and it is possible to derive corrections to the
velocity which yield the well known result $v - v_{\mathrm{min}}
\sim 1/\log^2 N$
\cite{derrida,pechenik}. 

The purpose of this paper is to extend this analysis of discrete
microscopic models to higher dimensions. In $d \geq 2$ the front
position is given by an interface at $x_{\|} = h(\xbot,t)$ which
separates the stable from the unstable domain. Due to the microscopic
fluctuations one expects the interface to roughen and its fluctuations
to be described asymptotically by the Kardar-Parisi-Zhang (KPZ)
\cite{kpzpaper}
equation, which features the simplest and most relevant non-linearity (in the
renormalization group sense) that considers local and lateral growth:
\beq\label{kpz}
\frac{\partial h}{\partial t} = \tilde{v} + \tilde{D} \nabla_{\bot}^2 h
+\frac{\lambda}{2} (\nabla_{\bot} h)^2 
+ \sqrt{2 \sigma}\ \eta_{\bot}(\xbot,t),
\eeq
where $\eta_{\bot}(\xbot,t)$ is an uncorrelated white noise, and
$\nabla_\bot$ is the divergence operator defined over the substrate
$\xbot$.  The interface is described by its mean position $\bar h(t)$
and its roughness,
\beq\label{roughness}
w^2(L_\bot,t) = \overline{\mean{[h(\xbot,t)-\bar h(t)]^2}},
\eeq
where $\mean{\cdots}$ means average over different realizations and
the bar denotes average over the substrate $\xbot \in
L_{\bot}^{\dbot}$ ($L_{\bot}$ is substrate lateral length). As it is well
known, in the KPZ equation $w$ obeys a scaling form $w(L_{\bot},t) =
t^\beta f(t/L_{\bot}^z)$, where $w \sim t^\beta$ if $t\ll L^z$ and $w
= w_{\mathrm{sat}}  \sim L^\alpha$ if $t \gg L^z$, where $\alpha =
\beta z$ is the roughness exponent \cite{barabasi}.

However, studies of microscopic realizations of Eq.\
(\ref{fisher_equation}) have questioned the applicability of
(\ref{kpz}) to describe the interface fluctuations in front
dynamics. Riordan {\em et al.} \cite{doering} studied the $A + A \rl
A$ scheme in various dimensions concluding that the interface
roughens in time for $\dbot \leq d_{\bot,c} = 2$ with $\beta_{\dbot =
1} = 0.27\pm 0.01$ and $\beta_{\dbot=2}\simeq 0$. These values
strongly differ from those of KPZ, which are $\beta_{\dbot =
1} = 1/3$ and $\beta_{\dbot=2} = 0.245$\cite{parisi}. For $\dbot >
d_{\bot,c}$, they observed that the interface is flat and described by
Eq.\ (\ref{fisher_equation}). However, the determination of these
values is hampered by the method used to obtain them, namely, using
the front width of the projected density $\rho(x_{\|},t)$ obtained
by integrating out all perpendicular dimensions $\xbot$.
%Moreover the density is calculated
%without taking into account the diffusive movement of the front
%position around its mean value and thus a final crossover to $w^2 \sim
%t$ is obtain. 
%To overcome these problems,
To overcome this problem, Tripathy and van Saarloos
\cite{saarloos1} studied the interface dynamics in $\dbot=1$ in terms of
the coarse-grained density of particles (see below) and obtained
$\beta_{\dbot=1} = 0.29\pm 0.01$, $\alpha_{\dbot=1} = 0.4\pm
0.02$. The authors interpreted this deviation from the KPZ exponents
($\alpha_{\dbot = 1}=0.5$) as an effect of the parallel dimension, which
cannot be integrated out to obtain an effective interface description
in $\dbot$ dimensions and thus, the exponents correspond to those of
the KPZ equation in $\dbot+1$ dimensions ($\alpha_{\dbot=2} =
0.393$) \cite{parisi}, although the interface is a surface
defined in a $\dbot$-dimensional substrate. More
recently\cite{saarloos2}, it has been conjectured that this situation
is more general and occurs in the dynamics of the so called ``pulled
fronts'' in which the region ahead of the front is important to
determine not only its deterministic properties (like the velocity)
but also the fluctuations of the effective interface dynamics
\cite{saarloos2}. Thus, it is important to check the validity of 
this conjecture for $\dbot > 1$. Although our results are consistent with 
those in \cite{saarloos1}, a closer analysis indicates that there is 
a crossover to the asymptotic behavior of KPZ equation in $d_{\bot}$
dimensions.

\begin{figure}
\begin{center}
\epsfig{file=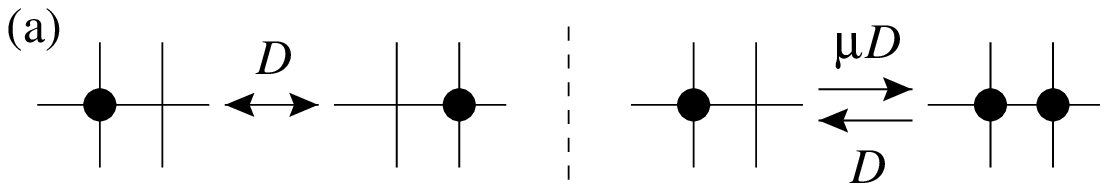, width=0.52in, angle=-90}\\
\epsfig{file=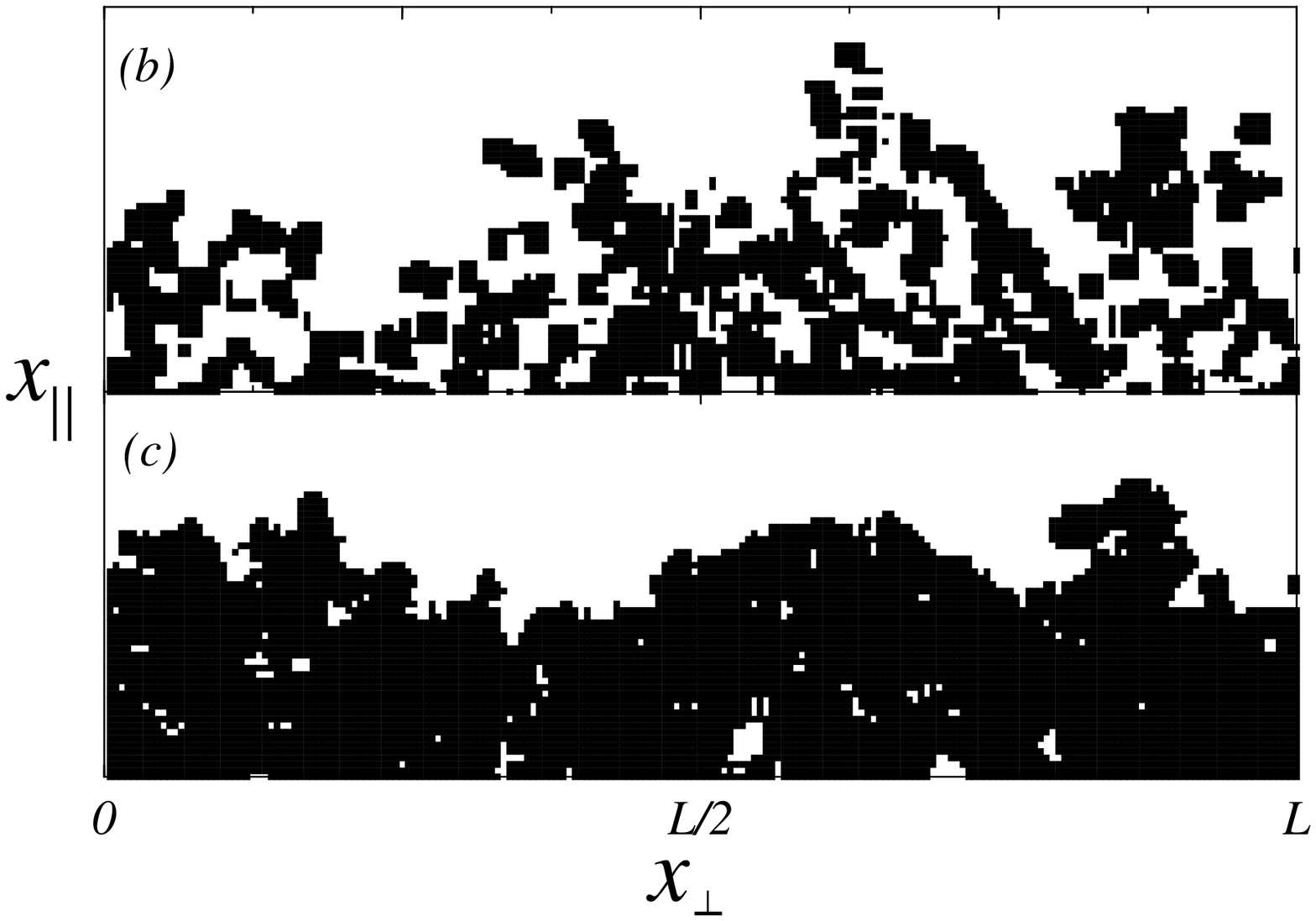,width=1.65in,angle=-90}
\caption{\label{trio} 
(a) Microscopic rules for our model: diffusion (left) and reaction
(right). The model is isotropic, i.e., all movements are equally
probable in any spatial direction. (b) and (c) Snapshots of the stable
domain for $\mu = 0.1$ (b), $\mu = 0.5$ (c) and $l = 5$.}
\end{center}
\vspace{-0.5cm}
\end{figure}

Our simulations are done on a $L_{\bot}^{\dbot}\times L_{\|}$ lattice
with $L_{\|} \gg L_{\bot}$. At each point of the lattice we define the
occupancy number $n(\rb,t)$ to be $0,1$ depending whether the site is
occupied by a particle or not. At each time step a particle is chosen
and can perform any of the processes in Fig.\ \ref{trio}a which, in
the mean-field approximation, yield $k_1 = \mu$ and $k_2 = 1 + \mu$,
i.e.  $\rho_{\mathrm{eq}}=\mu/(1+\mu)$ \cite{doering}. Diffusion rate
is chosen to be $D=1/(2d)$. Initial conditions are set up by
distributing particles uniformly over a domain $L_{\bot}^{\dbot}\times
L_{\|}^{(i)}$ with density $\rho_{\mathrm{eq}}$ and $L_{\|}^{(i)}$ of
the order of hundreds of lattice spacings. The interface is determined
using a local coarse-grained density $\rho_l(\rb,t)= l^{-d}\sum_{\sbo
\in D_l} n(\sbo,t)$ where $D_l$ is a squared domain of lateral size
$l$ around the $\rb$ site. The stable domain of the front is defined
by those sites for which $\rho_l(\rb,t) >
\frac12\rho_{\mathrm{eq}}$ and the position of the interface
$h(\xbot,t)$ is the largest value of $x_{\|}$ inside the stable domain
for fixed $\xbot$
\cite{saarloos1}.

\begin{figure}
\begin{center}
\epsfig{file=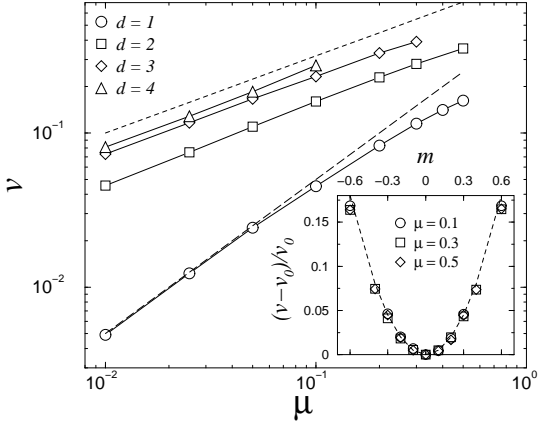, width=2.3in, angle=-90}
\caption{Front velocity for different
dimensions. Dashed lines are mean-field prediction $v = 2(D\mu)^{1/2}$
and $d=1$ exact solution $v = D\mu$. Values for $d \ge 2$ are divided
by $2D^{1/2}$. Error bars are smaller than the symbol size. Inset:
renormalized velocity as a function of the tilt $m$ for $\dbot =1$.
Dashed line is the parabola $m^2/2$ and $v_0=v(m=0)$.}\label{vel}
\end{center}\vspace{-0.5cm}
\end{figure}

Simulations show that the interface advances linearly in time, $\bar h
= v t$, where from we can extract the velocity of the front in any
dimension. Values obtained are reported in Fig.\ \ref{vel} and
compared with the deterministic prediction $v= v_{\mathrm{min}} =
2\sqrt{D\mu}$ of Eq.\ (\ref{fisher_equation}). We recall that in $d=1$
the stochastic model admits an exact solution \cite{doe_jsp}, and the
front advances with velocity $v = D \mu$. The difference between both
expressions reflects the importance of fluctuations when $N=1$ and
$d=1$, and agrees with the breakdown of the approximation
$v-v_{\mathrm{min}} \sim \log^{-2} N$ when $N=1$. Interestingly, for
$d > 1$ we recover the law $v \sim \mu^{1/2}$, although the value is
still corrected due to the fluctuations in the number of particles.
Nevertheless, the velocity change is smaller when the dimension is
increased, as fluctuations get averaged over the transverse
dimensions. Moreover, assuming that Eq.\ (\ref{kpz}) holds,
we can obtain the value of $\lambda$ by tilting the substrate with an
overall slope $m$ and measuring the velocity. As expected we 
obtain $v(m) = v(0) + \frac12 v(0) m^2$, i.e. $\lambda = v(0)$
\cite{barabasi}.

\begin{figure}
\begin{center}
\epsfig{file=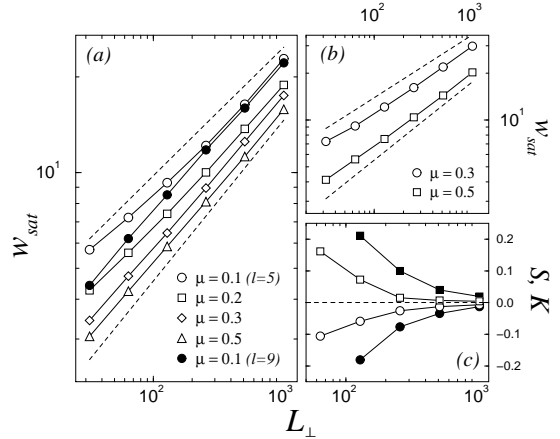, width=2.3in, angle=-90}
\caption{(a) Roughness as a function of $L_{\bot}$ for $d=2$. Dashed
lines correspond to power laws $L_{\bot}^\alpha$ with $\alpha = 0.4$
and $\alpha = 1/2$. (b) Same
as in (a) but for the directed percolation model with $c=0.1$. (c) Skewness
(circles) and kurtosis (squares) as a function of $L_{\bot}$ for $\mu
= 0.1$ (full symbols) and $\mu = 0.5$ (open symbols). Error bars are about
0.05 but have been omitted for clarity reasons. }\label{w2}
\end{center}\vspace{-0.5cm}
\end{figure}

For $d=2$ the interface roughens subject to internal fluctuations, and
the roughness grows like $w(t) \sim t^\beta$ until it saturates to a
constant value $w_{\mathrm{sat}} \sim L_{\bot}^\alpha$ (see Fig.\ \ref{w2}).
When $\mu = 0.1$ we obtain $\beta_{\dbot = 1} = 0.27 \pm 0.01$ and
$\alpha = 0.40\pm 0.01$, consistent with
\cite{saarloos1}. Nevertheless, a closer inspection of the data
suggests that there is a crossover from non-Gaussian fluctuations at
small scales to Gaussian ones at large scales. Although this crossover
could be guessed, for instance, from the value of $w_{sat}(L_\bot)$ for
$\mu = 0.1$ (which is seen to change its slope in a log-log plot), it
becomes clearer if we look at higher moments of the $h(\xbot,t)$
distribution, e.g. the skewness, $\scal =
\overline{\mean{(h(\xbot,t)-\bar h)^3}}/w^{3/2}(t)$ and the kurtosis,
${\cal K}=\overline{\mean{(h(\xbot,t)-\bar h)^4}}/w^2(t)-3$ when $t \to
\infty$ as seen in Fig.\ \ref{w2}. Both
approach the Gaussian asymptotic regime given by Eq.\ (\ref{kpz}) (i.e.
$\scal = \kcal = 0$) when $L_{\bot} \to \infty$. Non-Gaussian
fluctuations in the system are due to the vague definition of the 
interface for small values of $\mu$ and $l$. In Fig.\ \ref{trio}b,c we
show different snapshots of the stable domain calculated through the 
coarse-grained density, $\rho_l(\rb)$. It is apparent that
when the number of particles inside the domain $D_l$ is small, the 
definition of the interface is vague and implies large steps (or large
slopes) which give non-Gaussian fluctuations in $h(\rb,t)$ at small
length scales. Nevertheless these {\em intrinsic fluctuations} of 
the interface are bounded and restricted to small length scales. The 
measured roughness is naturally decomposed into contributions due
to intrinsic fluctuations, $w_i$, and long-wavelength fluctuations $w_0$
\cite{wolf,tammaro}
\begin{equation}\label{roughness2}
w^2(L,t) = w_i^2 + w^2_{0}(L,t),
\end{equation}
where $w_i$ is a non-universal constant that depends on the number of
particles, $n_{l}$, inside the domain $D_l$ used to determine the
interface. Direct fit of expression (\ref{roughness2}) to a power law
when $t \to \infty$ gives lower values of the exponent $\alpha$, while
we recover $\alpha = 1/2$ when fitting the data in Fig.\ \ref{w2}
using expression (\ref{roughness2}) and assuming that $w_0 \sim
L^\alpha$, showing that long-wavelength fluctuations are indeed
described by the KPZ equation in $\dbot$ dimensions.  In order to
reduce the effect of the intrinsic width, we can enlarge the size of
the domain and/or the value of $\mu$, as we can see in Fig.\ \ref{w2},
where a direct fit to a power law of the data for $\mu = 0.1$ and
$l=9$ gives $\alpha = 0.47\pm 0.01$. In this sense, $n_l$ plays the
role of a {\em noise-reduction parameter}. The situation then is
completely reminiscent of what happens in the Eden model \cite{wolf}
and has been also observed in many other RD systems
\cite{tammaro}.

\begin{figure}
\begin{center}
\epsfig{file=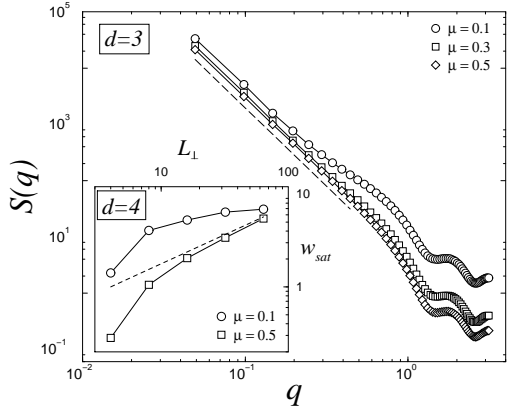, width=2.3in, angle=-90}
\caption{Structure factor as a function of the wave number $q$ for
$d = 3$ and $L_{\bot} = 128$. Dashed line correspond to the KPZ
prediction in $\dbot = 2$, $S(q)\sim q^{-2+2\alpha}$ with $\alpha = 0.393$.
Inset: Saturation roughness as a function of $L_{\bot}$ in
$d=4$. Dashed line corresponds to the power law $L_{\bot}^\alpha$ with
$\alpha = 0.313$.}\label{sdeq3d}
\end{center}\vspace{-0.5cm}
\end{figure}

In higher dimensions results are analogous. In $d = 3$ we do observe a
significant increase in the roughness with $L_{\bot}$, contrary to
what was observed in \cite{doering}. The reason for this discrepancy
is that the roughness measured in
\cite{doering} is also a linear combination of the long-wavelength
roughness $w_0(L_{\bot},t)$ and the front width, $\xi$. In $d=2$ we have
$w_0(L_{\bot},t) > \xi$, while in $d=3$, $w_0(L_{\bot},t)
\simeq \xi$ (for the simulated values of $L_{\bot}$), and this is
responsible for an apparent value of $\alpha_{\dbot=2} \simeq
0$. Actually, using the structure factor $S(\bqbot,t) = \langle\hat
h(\bqbot,t) \hat h(-\bqbot,t)\rangle$ where $\hat h(\bqbot,t)$ is the
Fourier transform of $h(\xbot,t)$, we observe in Fig.\ \ref{sdeq3d} a
crossover to the asymptotic scaling $S(\bqbot \ll 1,t\to\infty) \sim
q^{-d-2\alpha}$ with $\alpha_{\dbot=2} = 0.393$ of KPZ equation
[oscillations for $q>1$ are due to artificial correlations in
$h(\rb,t)$ introduced by the coarse-grained density]. Finally, in
$d=4$ we observe two different behaviors: for small values of $\mu$
the roughness seems to saturate to a value independent of $L_{\bot}$
while for $\mu = 0.5$ we recover the asymptotic prediction of Eq.\
(\ref{kpz}), $\alpha_{\dbot=3} = 0.313$ \cite{parisi}. This difference
may be due to the existence of a phase transition for the KPZ equation
in $\dbot > 2$, in which, depending on the control parameter
$g^2=\lambda^2 \sigma D^{-3}$, the interface is flat (for $g<g_c\simeq
6.28$) or rough (for $g>g_c$). Our results suggest that the interface is
flat for $\mu=0.1$ (i.e. $w^2$ does not scale with $L_{\bot}$) while it
is rough for $\mu=0.5$, although more simulations are needed.

We have also performed simulations for the $N=1$ {\em directed
percolation} model in which the decay $A \to \emptyset$ is allowed
with rate $c$. The model presents a non-trivial phase transition for
finite values of $\mu$ and $c$ in which the stable state changes from
an active ($\rho_{\mathrm{eq}}\neq 0$) to an absorbing state
($\rho_{\mathrm{eq}} = 0$) for $d \leq 4$ \cite{ref_dp}. In this case,
the mesoscopic dynamics of the system are given by \cite{cardy,rmp}
\begin{equation}\label{dirperc}
\frac{\partial \rho}{\partial
  t} = D \nabla^2 \rho + k_1 \rho -k_2 \rho^2 +
  \sqrt{k_3 \rho}\ \eta(\rb,t),
\end{equation}
the only difference with (\ref{langevin}) being the form of the noise
term.  Away from the transition point (which occurs when $\mu=0.22\pm
0.01$ for fixed value of $c=0.1$), interface fluctuations behave
similarly to the previous model (see Fig.\ \ref{w2}b), confirming that
our results do not depend on the specific choice of the microscopic
rules, as long as the mean-field limit of them coincides with
(\ref{fisher_equation}).

In summary we have studied two microscopic stochastic realizations
of Fisher's equation with $N=1$. It has been shown that with a
proper definition of the interface, long-wavelength fluctuations
are fully described within the KPZ scaling in $\dbot$ substrate
dimensions. Thus, for $d \leq 4$ there is not an upper critical
dimension above which (\ref{fisher_equation}) is valid. Only for $d
= 4$ does the interface seem to undergo a phase transition between a
flat interface [which can now be explained by (\ref{fisher_equation})]
and a rough phase, depending on the microscopic parameters.

Finally, we comment on the possibility to observe the conjecture in
\cite{saarloos2} using microscopic RD systems. It is possible to argue
that the results in this paper deviate from those of the conjecture
because the limit $N \to \infty$ in which Eq.\
(\ref{fisher_equation}) is valid has not been reached. Thus for large
values of $N < \infty$ \cite{porque} we should be able to observe a
crossover to the universality class proposed in
\cite{saarloos2}. However, the conjecture requires that the noise
term scales as $\rho \eta(\rb,t)$ when $\rho \to 0$, which we call {\em
multiplicative noise}. This type of noise can be obtained, for
instance, assuming that the reaction rates fluctuate in time due to
environmental variations (external noise). As argued in \cite{munoz},
{\em it is impossible to find such an internal noise in a microscopic
RD model}. For example, in the models analyzed here, the noise term
scales as $\rho^{1/2}\eta(\rb,t)$ for $\rho \to 0$ [see Eqs.\
(\ref{langevin}) and (\ref{dirperc})]. Thus, even if the parallel
dimension is relevant to determine the interface properties of the
front for large values of $N$, we expect internal fluctuations to give
a different universality class from the one proposed in
\cite{saarloos2}.

We thank J. L. Cardy for illuminating discussions and 
D. B. Abraham, A. Buhot, R. Cuerno, J. P. Garrahan, W. van
Saarloos and G. Tripathy for useful comments. This work has
been supported by EPSRC Grant No. GR/M04426 and EU Grant
No. HPMF-CT-2000-00487.

{\em Note added}: After completion of this paper we became aware 
of a recent preprint, R. A. Blythe and M. R. Evans, cond-mat/0104393, in 
which a similar crossover is found for a model in the directed percolation
universality class of (\ref{dirperc}) in $\dbot = 1$.

%\bibitem{wolf} D. E. Wolf and J. Kert\'esz, J. Phys. A {\bf 20}, L257
%  (1987); J. Kert\'esz and D. E. Wolf, {\em ibid.} {\bf 21}, 747
%  (1988).  
%\bibitem{avraham} D. ben-Avraham and S. Havlin, {\em Diffusion and
%    Reactions in Fractals and Disordered Systems} (Cambridge
%  University Press, Cambridge, 2001).
%\bibitem{parisi} E. Marinari, A. Pagnani, G. Parisi, J. Phys. A:
%Math. Gen. {\bf 33}, 8181 (2000).
%\bibitem{porque} Note that in the limit $N \to \infty$ the noise terms
%in (\ref{langevin}) and (\ref{dirperc}) vanish and hence the interface
%does not fluctuate.
%\bibitem{munoz} M. J. Howard and U. C. T\"auber, J. Phys. A: Math. Gen.
%{\bf 30} 7721 (1997); M. A. Mu\~noz, Phys. Rev. E {\bf 57} 1377 (1998).
%\end{thebibliography}


\begin{thebibliography}{99}
\bibitem{porque} Note that $N$ has to be finite, because in the 
limit $N \to \infty$, noise terms in (\ref{langevin}) and
(\ref{dirperc}) vanish and hence the interface does not fluctuate.
\end{thebibliography}
\end{document}